\documentclass[preprint,12pt]{elsarticle}

\usepackage{amssymb}
\usepackage{amsmath}
\usepackage{float}
\usepackage{geometry}
\journal{Nuclear Physics B}

\begin{document}

\begin{frontmatter}



\title{A 3D Generation Framework from Cross Modality to Parameterized Primitive} 


\author[a,b]{Yiming Liang\fnref{label1}} 

\author[a,b,c]{Huan Yu\fnref{label1}}
\fntext[label1]{The two authors contribute equally to this work.}

\author[a,b]{Zili Wang\corref{cor1}}
\cortext[cor1]{Zili Wang is the corresponding author}
\ead{ziliwang@zju.edu.cn}

\author[a,b]{Shuyou Zhang}

\author[a,b]{Guodong Yi}

\author[a,b,c]{Jin Wang}

\author[a,b]{Jianrong Tan}
\affiliation[a]{organization={Zhejiang University},
            addressline={State Key Laboratory of Fluid Power \& Mechatronic Systems}, 
            city={Hangzhou},
            postcode={310058}, 
            state={Zhejiang},
            country={China}}
    
\affiliation[b]{organization={Zhejiang University},
            addressline={School of Mechanical Engineering}, 
            city={Hangzhou},
            postcode={310058}, 
            state={Zhejiang},
            country={China}}

\affiliation[c]{organization={Robotics Research Center of Yuyao City},
            addressline={Yuyao}, 
            city={Ningbo},
            postcode={315400}, 
            state={Zhejiang},
            country={China}}

\begin{abstract}
Recent advancements in AI-driven 3D model generation have leveraged cross modality, yet generating models with smooth surfaces and minimizing storage overhead remain challenges. This paper introduces a novel multi-stage framework for generating 3D models composed of parameterized primitives, guided by textual and image inputs. In the framework, A model generation algorithm based on parameterized primitives, is proposed, which can identifies the shape features of the model constituent elements, and replace the elements with parameterized primitives with high quality surface. In addition, a corresponding model storage method is proposed, it can ensure the original surface quality of the model, while retaining only the parameters of parameterized primitives. Experiments on virtual scene dataset and real scene dataset demonstrate the effectiveness of our method, achieving a Chamfer Distance of $3.092\times10^{-3}$, a VIoU of 0.545, a F1-Score of 0.9139 and a NC of 0.8369, with primitive parameter files approximately 6KB in size. Our approach is particularly suitable for rapid prototyping of simple models.
\end{abstract}







\end{frontmatter}



\section{Introduction}
With the rapid development of artificial intelligence generation technology and computer graphics, three-dimensional model representation technology has become a popular research direction and has a wide range of applications in multiple fields, including but not limited to virtual reality, medical image processing, industrial design and manufacturing, game development, etc.~\cite{li2023generative,Zilingling2017Survey}. 

Traditional 3D model generation techniques typically rely on multi-view Primitive and complex geometric optimization algorithms. However, these methods~\cite{furukawa2010towards,schonberger2016pixelwise,choi20103d} face significant limitations when applied to complex scenes or objects with missing texture information. For instance, conventional approaches often require substantial prior knowledge and assumptions, such as object shape, lighting conditions, and texture details, which restrict their applicability in real-world scenarios. Furthermore, they usually necessitate multiple input images and involve computationally intensive processes, thereby hindering automation and reducing efficiency.

In contrast, the AI-based 3D model generation technology offer notable advantages. They can accept text descriptions and single images as inputs, and generate implicit 3D representations through multi-view depth image synthesis. By combining implicit representations~\cite{min2023tsdf} and sparse optimization~\cite{tatarchenko2017octree}, these methods can produce high-resolution outputs using limited memory, ultimately enabling the zero-shot generation of entirely novel 3D models.

Although the existing 3D model generation made remarkable progress and are capable of producing models that closely align with user requirements, the following challenges remain:

(1) High storage demand. Existing methods~\cite{lorensen2020history} typically generate 3D models by extracting explicit mesh representations from implicit 3D representation~\cite{mildenhall2021nerf,kerbl20233d}. For voxel representation, it is necessary to divide 3D space into a dense regular grid, and each voxel needs to store the occupancy state or numerical value. While high-resolution voxels can capture fine details, their memory requirements scale cubically with resolution. 
For instance, a $256^3$ voxel grid needs to store more than 16 million voxel information, with a memory footprint of up to 0.54GB~\cite{golyanik2018hdm}. 

(2) Model surface quality. The low surface quality of the model often limited by the resolution and topological structure constraints. Low resolution voxels (such as $32^3$) may lead to the loss of details~\cite{bednarik2018learning}. Mesh-based methods depend on the deformation of the initial template (such as an ellipsoid) and cannot flexibly handle multiple holes or complex topologies.

The existence of these problems has posed significant challenges to the advancement of 3D model generation technology. At present, there is a pressing requirement for a generation method that can improve the surface quality of the model while reducing the model storage requirements. 

\begin{figure*}[ht]
  \centering
  \includegraphics[width=\linewidth]{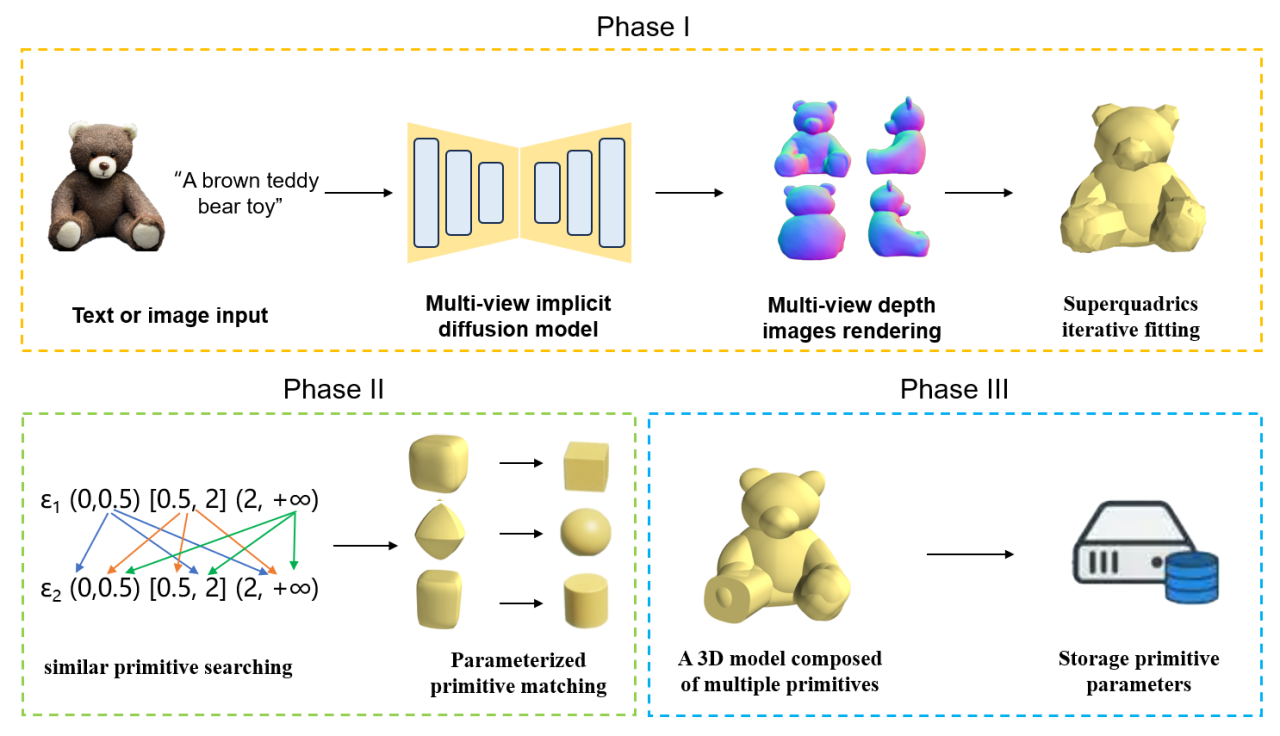}
  \caption{The framework of our method. In the first stage, an implicit diffusion model is introduced to synthesize multi-view depth images, and the target model is iteratively fitted with superquadrics. In the second stage, the parameterized primitive searching algorithm is executed to match the corresponding parameterized primitives for superquadric elements in the target model. In the third stage, use parameterized primitives to synthesize the target model and save the parameters of the model elements. }\label{fig1}
\end{figure*}

A zero-shot 3D basic model generation framework is proposed, to address the aforementioned challenges. The process of our method is shown in Fig~\ref{fig1}, which is divided into three stages. In the first stage, with text and images input as guidance conditions, multi-view depth images of the target model are generated using an implicit diffusion model, then, the truncated signed distance field is introduced for superquadric iterative fitting. In the second stage, searching for parameterized primitives that are similar to superquadric elements in the target model. In the third stage, execute primitive fitting and matching algorithm, replace superquadric elements with parameterized primitives, and store the target model composed of parameterized primitives. Our main contributions are summarized as follows:

(1) A primitive fitting and matching algorithm is proposed, which can replace the superquadric elements that make up the model with parameterized geometries with higher surface quality, thereby enhancing the overall quality of the 3D model.

(2) A 3D model storage method is proposed, which reduces the storage requirements of the model by retaining only the parameters of primitive elements.

(3) A three-stage 3D model generation method based on multimodal information is proposed. The method takes text and image information as inputs, and generates 3D models composed of parameterized Primitive under zero-shot condition.

\section{Related works}
\subsection{Implicit diffusion model}
Due to the lack of basic information on spatial geometric structures, single image based 3D reconstruction technology has long been a challenging problem in the field of computer vision. The early 3D model generation technology~\cite{wu2017marrnet,choy20163d,kar2017learning} was mainly based on supervised learning. The explicit representation method was used to represent 3D models as binary or real-valued 3D tensors. However, such methods require a large amount of supervised data, resulting in huge data acquisition costs. In recent years, the proposal of implicit diffusion models~\cite{ho2020denoising} and their application in text and image generation have provided new ideas for single-image-based 3D reconstruction. Some methods~\cite{wang2023imagedream,shi2023mvdream,long2023wonder3d,liu2023zero, abdullah2024vae} learn the complex structure of data distribution by gradually denoising training, score distillation sampling(SDS) technology and pre-trained 2D diffusion model are used to guide the optimization of 3D representation, so that these methods can use the prior knowledge of 2D diffusion model to generate a 3D model consistent with the prompts without any clear 3D data. However, these methods need to obtain the grid model through the marching cubes algorithm~\cite{lorensen2020history}, resulting in extremely high model storage requirements and difficulties in model calculation and rendering. 

\subsection{3D model of primitive synthesis}
Some researches have explored the method of synthesizing 3D models through basic elements, in order to reduce the storage requirements of 3D models and allow users to edit the details of the generated 3D models. Existing methods are mainly aimed at decomposing 3D models into multiple simple primitives to achieve shape characterization. In addition to simple primitives such as cuboids and ellipsoids, hyperellipsoids~\cite{kluger2021cuboids}, anisotropic Gaussian~\cite{genova2019learning} and convexes~\cite{fahim2022enhancing} are also proposed. Smirnov et al.~\cite{smirnov2020deep} used parametric and constructive solid geometric operations on the cube to decompose the 3D model into simple geometric primitives, introduced surface loss and normal alignment loss functions, and optimized the geometric consistency between the generated shape and the target shape. Genova et al.~\cite{genova2019learning} proposed a method based on structured implicit function to model the shape as a combination of a group of shape elements with local influence through implicit surface representation. Paschalidou et al.~\cite{paschalidou2021neural} learned the 3D shape primitive representation of expressiveness through a reversible neural network (INN). Saporta et al.~\cite{saporta2022unsupervised} used a recursive neural network architecture to reconstruct 3D models from geometric primitives in an unsupervised learning manner.Such methods rely on a large number of datasets to train the deep learning network, and need to consider specifying the number of fitting primitives and the target model category before training, which limits the range of shapes that can be fitted. Therefore, Liu et al.~\cite{liu2023marching} proposed a method called Marching-Primitives(MP). By iteratively fitting the truncated signed distance field of the model using a superquadric, each model can be analyzed and fitted separately, expanding the range of models that can be generated. However, the input of this method is limited by the existing 3D model, and the surface quality of the model needs to be improved.

\section{Methodology}

\subsection{Multi-view depth image synthesis and superquadric iterative fitting}

\subsubsection{Multi-view depth image synthesis based on implicit diffusion model}
We introduce the pretrained ImageDream~\cite{wang2023imagedream} as a model for multi-view depth image synthesis. Due to the lack of a module for extracting multi-view depth images in the model architecture, we first use ImageDream to generate multi-view images of the target model, and use score distillation sampling loss function~\cite{poole2022dreamfusion} to guide the optimization of neural radiation field, and then use the optimized neural radiation field to render depth images from different perspectives. Finally, we use the sampling method of NeRFStudio~\cite{tancik2023nerfstudio} to sample depth images from different perspectives from the optimized implicit neural radiation field. We set the number of depth images to 48, and six at each of the eight azimuth angles, to provide sufficient model reconstruction information for the subsequent truncated signed distance field (TSDF) aggregation.

TSDF is composed of the truncated signed distance of each spatial point. When calculating the TSDF value of point $P$ under the space coordinate system of neural radiation field, our method uses the camera's internal and external parameter matrix to calculate the mapping point of point $P$ after transformation of the camera coordinate system and the depth image coordinate system.

\subsubsection{Superquadric iterative fitting}
Since the Marching Cubes algorithm will cause a large storage requirement when extracting the mesh model from the truncated signed distance field, we will use the mesh model extraction algorithm based on Marching Primitive~\cite{liu2023marching} in this step to output the 3D model composed of multiple superquadrics. First, after inputting voxels $V$, a group of decreasing sign distance threshold sequence ${T^{\rm{c}}} = \{ t_1^c,t_2^c, \ldots ,t_m^c,t_{m + 1}^c\} $ is defined. Under the current set threshold, all voxels whose $TSDF$ value is less than the sign distance threshold are extracted, and based on the connectivity of 26 domains, they are divided into multiple disjoint connected regions $S_m$. Then, the connected regions whose number of voxels is less than the threshold $N_c$ are filtered out to obtain an effective superquadric candidate region $\bar{S}_m$. We set the minimum truncation symbol distance field value as the initial threshold value for connectivity calculation, and then use the weight $\alpha$ to attenuate the threshold value to ensure that there is an effective superquadric candidate region. This process can be expressed by the following formula:
\begin{align}\left\{\begin{aligned}
t_1^c&=\min _{\mathbf{x}_i \in \mathbf{V}} t\left(\mathbf{x}_i\right)\\ t_{m+1}^c&=\alpha t_m^c, m=1, 2, . . . \\
S_m&=\left\{\mathcal{S}_k, k=1, 2, \ldots, \left|S_m\right|\right\}\\
\bar{S}_m&=\left\{\mathcal{S}_k \in S_m, \left|\mathcal{S}_k\right| \geq N_c\right\} \subseteq S_m
\end{aligned}\right. \end{align}

\noindent
where $\alpha$ is set to 0.6 in order to extract as many small detailed spatial structures as possible. The minimum voxel threshold $N_c$ in the isosurface is 5.

Then the effective candidate regions are fitted by superquadric iteration. superquadric has the following parameters: $\theta=(\varepsilon_1, \varepsilon_2, \mathbf{T}, \mathbf{R}, \mathbf{S})$, Where $\varepsilon_1$ and $\varepsilon_2$ are shape parameters, $\varepsilon_2$ determines the shape of the superquadric in the $xy$ plane, $\varepsilon_1$ determines the shape of the superquadric in the $z$ direction, $\mathbf{T} \in \mathbb{R}^3$ is the translation vector, $\mathbf{R} \in \mathbb{R}^3$ is the rotation vector represented by the Euler angle, and $\mathbf{S} = (a, b, c)\in \mathbb{R}^3$ is the size parameter, then the superquadric can be defined by the following implicit equation:

\begin{equation}
f\left( x \right) = {\left( {{{\left( {{x \mathord{\left/
 {\vphantom {x a}} \right.
 \kern-\nulldelimiterspace} a}} \right)}^{{2 \mathord{\left/
 {\vphantom {2 {{\varepsilon _2}}}} \right.
 \kern-\nulldelimiterspace} {{\varepsilon _2}}}}} + {{\left( {{y \mathord{\left/
 {\vphantom {y b}} \right.
 \kern-\nulldelimiterspace} b}} \right)}^{{2 \mathord{\left/
 {\vphantom {2 {{\varepsilon _2}}}} \right.
 \kern-\nulldelimiterspace} {{\varepsilon _2}}}}}} \right)^{{{{\varepsilon _2}} \mathord{\left/
 {\vphantom {{{\varepsilon _2}} {{\varepsilon _1}}}} \right.
 \kern-\nulldelimiterspace} {{\varepsilon _1}}}}} + {\left( {{z \mathord{\left/
 {\vphantom {z c}} \right.
 \kern-\nulldelimiterspace} c}} \right)^{{2 \mathord{\left/
 {\vphantom {2 {{\varepsilon _1}}}} \right.
 \kern-\nulldelimiterspace} {{\varepsilon _1}}}}} = 1
\label{superquadric}
\end{equation}

\noindent

Subsequently, the truncated signed distance field of this superquadric $\Theta_k$ is also calculated, and the characteristic vector of the superquadric is gradually optimized by minimizing the value. The superquadric fitting the target shape is obtained. The optimization formula is:
\begin{equation}
    \begin{aligned}
    \boldsymbol{\Theta}_k=\underset{\boldsymbol{\Theta}_k}{\arg \min } \sum_{\mathbf{x}_i \in \mathbf{V}} W_{ik}\left\|t_{\boldsymbol{\theta}_k}\left(\mathbf{x}_i\right)-t\left(\mathbf{x}_i\right)\right\|_2^2
    \end{aligned}
 \end{equation}

\noindent
Where the $W_{ik}$ is the relevant weight between the superquadric and the target voxel, and $t(.)$ represents the truncated signed distance field value of the target voxel. The detailed derivation process of this formula can be seen in \cite{liu2023marching}. Since it is not the key point of this article, it will not be described in detail.

\subsection{Similar parameterized primitives searching}
Considering that users prefer to use simple and reusable cubes, ellipsoids or other simple geometric primitives when generating zero-shot 3D basic models, a similar parameterized primitive searching method is proposed in phase II, which takes the 3D model after superquadric iteration fitting as the input, and outputs parameterized primitive elements similar to superquadric elements, to meet the user's usage tendency.

\begin{figure}[!b]
  \centering
  \includegraphics[width=0.5\linewidth]{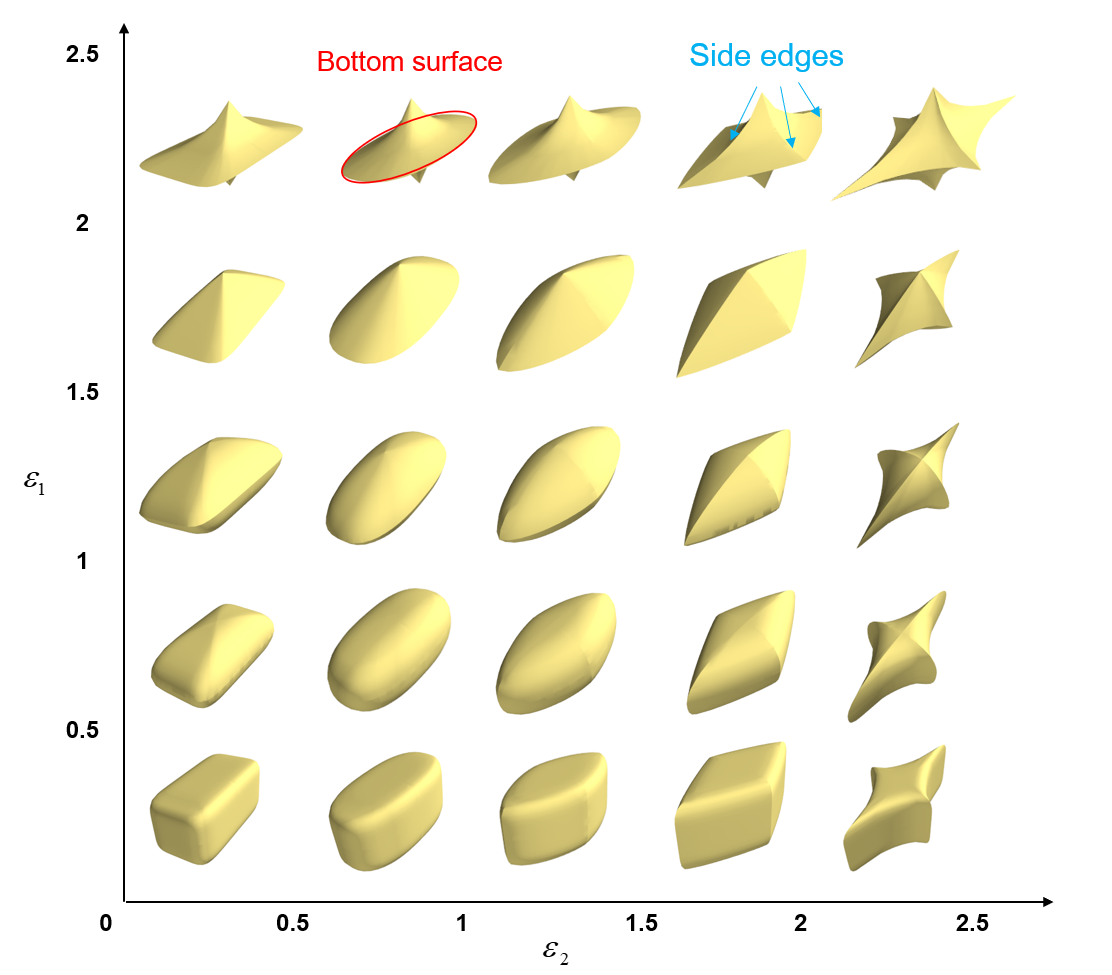}
  \caption{Shape change law of superquadrics. As $\varepsilon_1$ and $\varepsilon_2$ change, the shape of the superquadric also changes in the $z$-direction and $xy$ plane. The bottom and side edges defined in the text are shown in the figure}\label{figreg}
\end{figure}

It can be seen from equation~\ref{superquadric} that the shape of the superquadric is determined by two parameters, $\varepsilon_1$ and $\varepsilon_2$. We set the size parameter $\mathbf{S} = (a, b, c)\in \mathbb{R}^3$ of the superquadric as 1, 2, and 1 respectively, so as to observe the change of the shape of the superquadric in the $x$, $y$, and $z$ directions. No rotation and displacement transformation is carried out on the superquadric, so that the center of the superquadric is at the origin, and both are symmetrical about the three coordinate axis. We studied the influence of $\varepsilon_1$ and $\varepsilon_2$ on the shape characteristics of the superquadric in different value ranges, and summarized the shape change law of the superquadric is shown in Fig~\ref{figreg}. In the figure, we refer to the plane of the superquadric on $xy$ as the bottom surface, and the line intersecting the base as the side edge. Therefore, the superquadric will have four side edges on each of its upper and lower surfaces. 

It can be seen from the figure that with the increase of $\varepsilon_1$, the superquadric gradually becomes sharp in the $z$ direction. When the value range of $\varepsilon_1$ is $(0,0.5)$, all the side edges of the superquadric are almost perpendicular to the $xy$ plane, exhibiting the characteristics of a cylinder in the $z$ direction; When the value range of $\varepsilon_1$ is $[0.5,2]$, the side edges away from the $xy$ plane and intersect at a point, exhibiting the characteristics of a cone in the $z$ direction; When the value range of $\varepsilon_1$ is $\left( {2, +\infty } \right)$, the side edges bend towards the $xy$ plane and intersect at a point, exhibiting the characteristics of a star shape in the $z$ direction. Therefore, the superquadric has three shape features in the $z$ direction.

Similarly, as $\varepsilon_2$ increases, the superquadric gradually becomes sharper on the $xy$ plane. When the value range of $\varepsilon_2$ is $(0,0.5)$, the bottom surface of the superquadric exhibits rectangular characteristics; When the value range of $\varepsilon_2$ is $[0.5,2]$, the bottom surface of the superquadric exhibits elliptical characteristics; when the value range of $\varepsilon_2$ is $\varepsilon_1$ is $\left( {2, +\infty } \right)$, the bottom surface of the superquadric exhibits star-like characteristics. So the superquadric also has three shape characteristics on the xy plane.

In summary, we divide $\varepsilon_1$ and $\varepsilon_2$ into three value intervals: $(0,0.5)$, $[0.5,2]$, and $\left( {2, +\infty } \right)$. Each representing a shape feature of the superquadric in the $z$-direction or $xy$ plane. Therefore, nine kinds of superquadrics with different shapes can be formed by combining a shape features in $z$ direction and a shape features in $xy$ plane. When receiving the superquadric element in the target model, this method will determine which interval the shape parameters of element should be located in, thereby determining the shape characteristics represented by the superquadric and determining the parameterized primitive that is similar to it.

\subsection{primitive fitting and matching algorithm}
\label{sec1}
\begin{figure*}[h]
  \centering
  \includegraphics[width=1\linewidth]{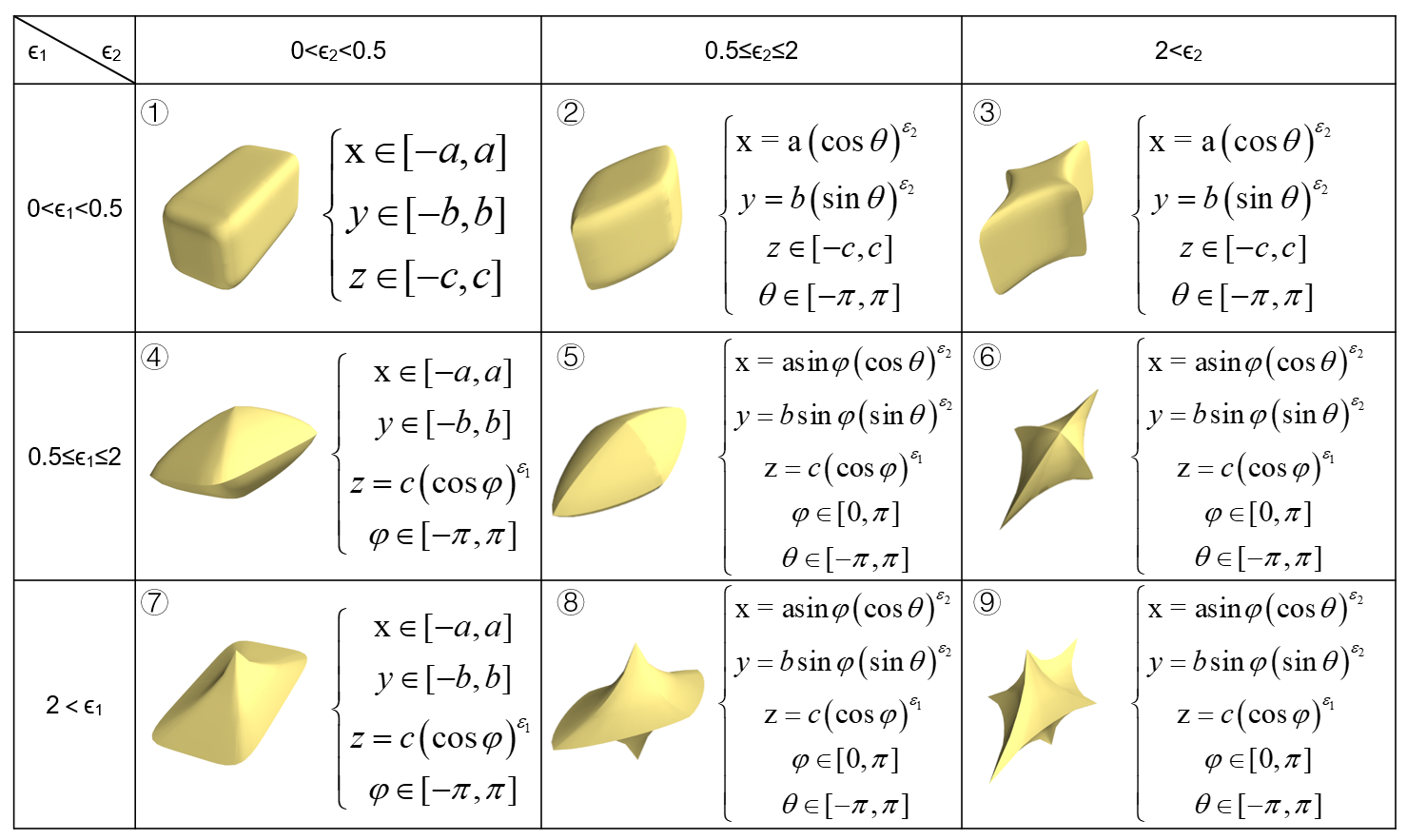}
  \caption{Nine types of superquadrics and their similar parameterized primitive expressions. The centers of the superquadrics in the figure are all at the origin, and the maximum values in the x-axis, y-axis, and z-axis directions are a, b and c, respectively}\label{figequ}
\end{figure*}

After finding similar parameterized primitives for the nine kinds of superquadrics, we will use polar coordinate equations to represent these parameterized primitives. In $z$ direction, the shape characteristics of cylinder, ellipsoid and star are represented by polar coordinate equations of cylindrical coordinate system, spherical coordinate system and star line respectively. On the $xy$ plane, the shape characteristics of the rectangular bottom, the elliptical bottom and the star bottom are respectively represented by the polar coordinate equations of similar figures. In addition, because the shape characteristics of the superquadrics change continuously with $\varepsilon_1$ and $\varepsilon_2$, the influence of these two shape parameters of the superquadrics needs to be considered in the representation equation of the parameterized primitive. To sum up, the nine types of superquadrics and their corresponding parameterized primitive representation equations are shown in Fig~\ref{figequ}. 

Combining the rotation vector $\mathbf{R}$ and the translation vector $\mathbf{T}$ of the superquadrics, as well as the polar coordinate equations of the parameterized primitives, our method will perform translation and rotation transformations, then performs the optimized fitting and matching of the target 3D model from parameterized primitives. 

For model storage, our method only retains the parameters (size parameters $\mathbf{S}=(a, b, c)$, shape parameters $\varepsilon_1$ and $\varepsilon_2$, translation vector $\mathbf{T}$, rotation vector $\mathbf{R}$) of all primitive elements in the model to reduce storage capacity. When the model needs to be used, simply read the stored parameters to reconstruct the target model.

\section{Experimental results and discussion}

\subsection{Dataset construction and evaluation indicators}\label{sec2}

We will prepare a virtual scene dataset and a real scene dataset for experimentation, to fully validate the generalization of the method. The virtual scene dataset is mainly composed of ShapeNet~\cite{chang2015shapenet}, the ShapeNet dataset is widely used in 3D model generation research, including more than 3000 object categories and 220000 models, which is suitable for component segmentation of simple models. In addition, we also selected test images and texts from multiple 3D generation models such as ImageDream~\cite{wang2023imagedream}, One-2-3-45++~\cite{liu2024one}, Wonder3D~\cite{long2023wonder3d}, MVDream~\cite{shi2023mvdream}, and TripoSR\cite{tochilkin2024triposr}, etc. to form the virtual scene dataset, in order to verify the model generation effect under text input conditions. The real scene dataset is mainly composed of CO3D~\cite{reizenstein2021common}, which provides rich real-world 3D data suitable for tasks such as 3D model reconstruction. In addition, it also includes some images from AKB-48~\cite{liu2022akb} and OmniObject 3D~\cite{wu2023omniobject3d}. 

    
For the effect evaluation of primitive fitting and matching, we select four commonly used indicators, Chamfer Distance (CD) Volumetric Intersection over Union (VIoU), F1-Score and Normal Consistency(NC). CD is a distance measurement method used to measure the similarity between two point clouds. Its basic idea is to compare the total distance of the nearest neighbor in two point sets. VIoU is an extension of IoU in 3D space, which is used to evaluate the overlap degree of 3D models. F1-Score takes into account both surface reconstruction accuracy and recall, and can evaluate the matching degree between the generated model and the reference model. NC evaluates the consistency between the surface normal vectors of the model and the reference model, and is suitable for assessing the fidelity of surface quality and geometric details, especially for comparing the reconstructed model with the reference model. These four indicators can fully assess the similarity between the fitted shape and the generated shape. The combination of these four indicators comprehensively reflects the quality effect of the primitives combination model based on text-image generation implemented by our algorithm.

\subsection{Experimental deployment environment}

All our experiments were conducted in a Windows 11 environment with an AMD Ryzen 7 9700X CPU @ 3.80GHz and an NVIDIA GeForce RTX 5060Ti. All the code was implemented based on Python 3.10. In terms of parameter settings. For TSDF, We set the voxel space size to $[-1^3, 1^3]$, and conduct 100 uniform samples for each dimension. A total of $10^6$ voxels are sampled. The truncation value is set to 1.2 times the voxel size. Finally, we set the grid resolution of the generated model to 100. Other parameters use default values. 

\subsection{Experimental results display}


\subsubsection{Validation experiment of primitive fitting effect}

In this section, we will compare our method with some state-of-the-art 3D model generation methods~\cite{liu2022robust,liu2023marching,fedele2025superdec} with virtual scene dataset and real scene dataset to verify that our method can generate 3D models that meet the input conditions of multimodel information. The introduction of datasets used in validation experiment have been mentioned in Section~\ref{sec2}.

\subsubsection{Validation experiment of primitive fitting effect on virtual scene dataset}
The quantitative experiment results on the virtual scene dataset is shown in Tab~\ref{tab1}. It can be observed that our method can achieve better fitting optimization and matching through a variety of simple primitives. A CD decrease of $3.092\times10^{-3}$, a VIoU improvement of 0.545, a F1-score improvement of 0.9139 and a F1-score improvement of 0.8369 are achieved. Compared with MP, our method reduced the CD by 37.6\% , increased the VIoU by 39.7\%, the F1-Score by 11.5\%, and the NC by 14.9\%,. It shows that our method does not cause degradation of fitting effect when replacing the superquadrics with simple primitives, and can generate a 3D model closer to the target model.  
      
    \begin{table}[H]
    \centering
    \caption{Quantitative experimental results on the virtual scene dataset}
    \label{tab1}
    \begin{tabular}{ccccc}
    \hline
        & EMS~\cite{liu2022robust} & SuperDec~\cite{fedele2025superdec}    & MP~\cite{liu2023marching}    & our method \\ \hline
    CD($\times10^{-3}$)$\downarrow$ & 13.1 & 6.38 & 4.95 & \textbf{3.09}      \\
    VIoU$\uparrow$     & 0.218 & 0.246 & 0.390 & \textbf{0.545}      \\ 
    F1-Score$\uparrow$     & 0.8572 & 0.8629 & 0.8193   & \textbf{0.9139}      \\ 
    NC$\uparrow$     & 0.6607 & 0.7101 & 0.7284 & \textbf{0.8369}      \\ \hline
    \end{tabular}
    \end{table}

\begin{figure}[h]
  \centering
  \includegraphics[width=\linewidth]{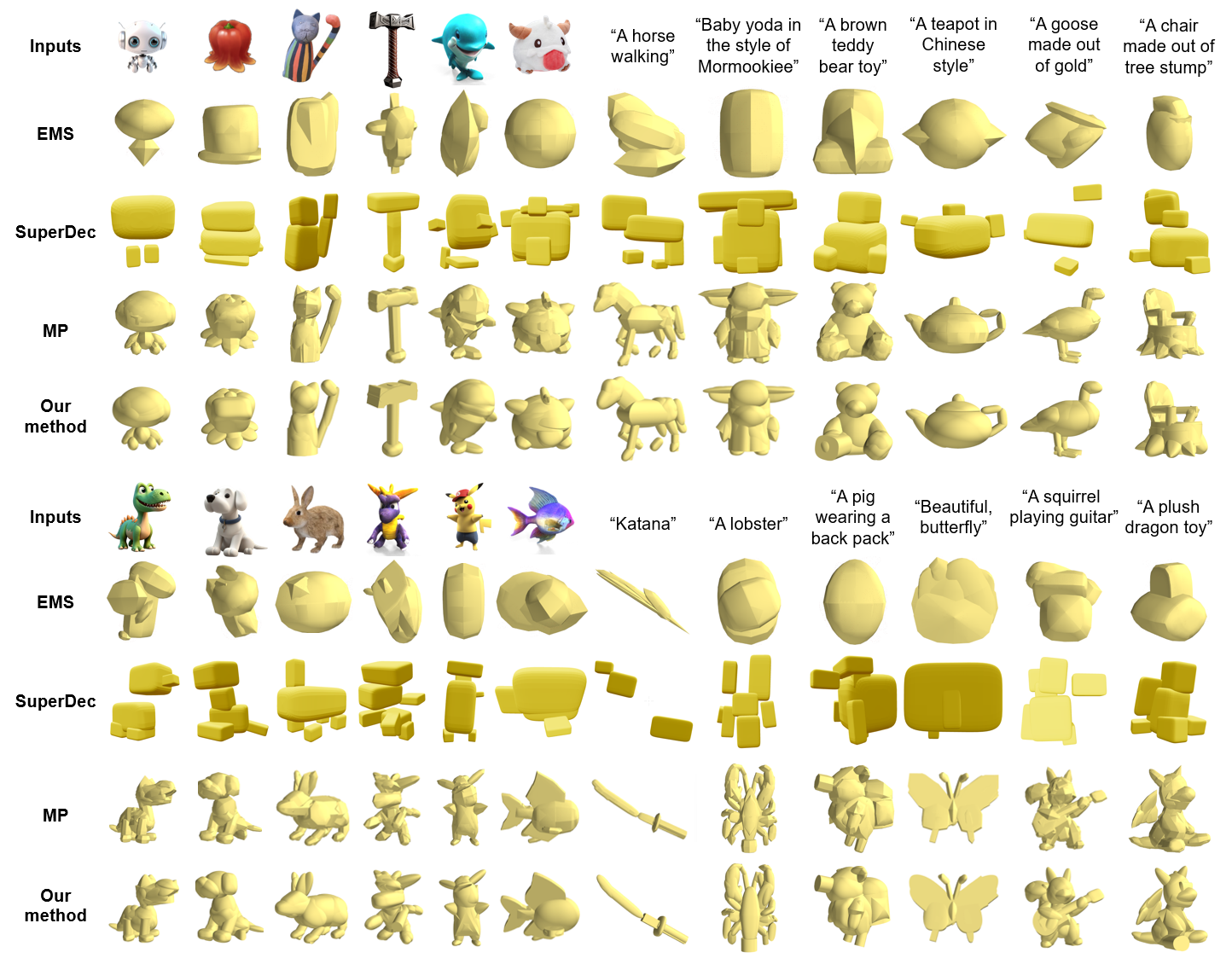}
  \caption{Qualitative experimental results based on image and text inputs. The first six columns are based on image input, while the remaining columns are based on text input}\label{fig_vir}
\end{figure}

\begin{figure}[h]
  \centering
  \includegraphics[width=1\linewidth]{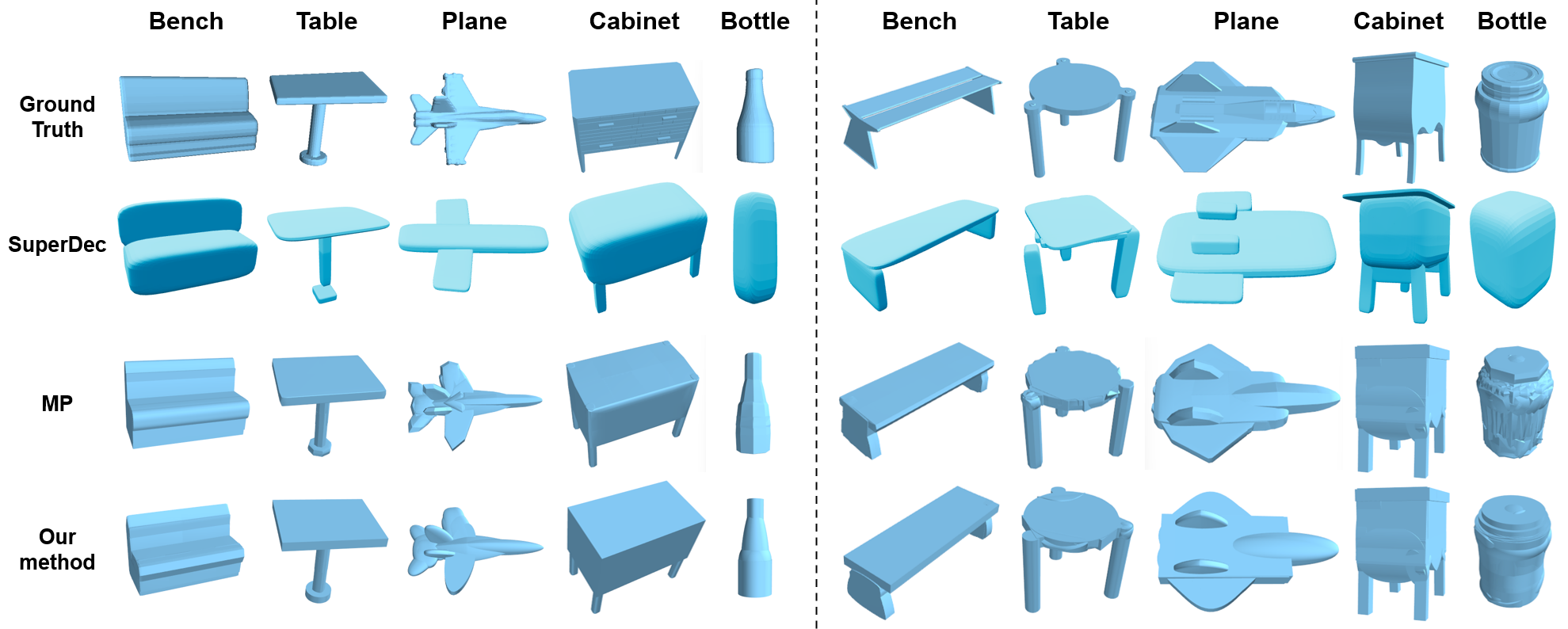}
  \caption{Qualitative experimental results on the ShapeNet dataset. We presented the test sample results for the categories of bench, table, plane, cabinet, and bottle in ShapeNet}\label{fig_sha}
\end{figure}

Our method is committed to receiving image and text inputs to achieve cross-modal 3D model generation. Therefore, in the qualitative experiment on the virtual scene dataset, we verified the 3D model generation effect generated by different input conditions. The experimental results based on image and text input conditions are shown in Fig~\ref{fig_vir}. It can be seen from the figure that our method can generate 3D models that meet the zero-shot input conditions, and our method uses a variety of simple primitives to optimize the fitting and matching of superquadrics elements, so that the surface quality of the target model is improved, more in line with the subjective aesthetics, and can also use the shape of primitives to repair some situations with poor generation effects.

\begin{table*}[h]
\centering
\caption{Quantitative results on the ShapeNet dataset}
\label{tab_sha}
\begin{tabular}{cccccccc}
\hline
& \multicolumn{3}{c}{CD($\times10^{-3}$)$\downarrow$}     & \multicolumn{3}{c}{VIoU$\uparrow$}           \\ 
   & SuperDec~\cite{fedele2025superdec} & MP~\cite{liu2023marching}    & our method     & SuperDec~\cite{fedele2025superdec}   & MP~\cite{liu2023marching}    & our method \\ \hline
bench   & 3.21  & 2.45  & \textbf{0.433}      & 0.381  & 0.433 & \textbf{0.798}     \\ 
table   & 1.76 & 2.17  & \textbf{0.330}       & 0.616 & 0.413 & \textbf{0.685}      \\ 
plane   & 2.38 & 0.840 & \textbf{0.293}       & 0.399 & 0.636 & \textbf{0.772}      \\ 
cabinet & 3.13 & 3.11  & \textbf{0.849}       & 0.403 & 0.373 & \textbf{0.753}        \\
bottle  & 5.49  & 2.16  & \textbf{0.760}       & 0.253  & 0.503 & \textbf{0.762}      \\ 
rifle   & 2.58  & 1.04  & \textbf{0.352}       & 0.531 & 0.567 & \textbf{0.682}      \\ \hline
mean    & 3.09 & 1.96  & \textbf{0.503}       & 0.431 & 0.488 & \textbf{0.742}      \\ \hline
& \multicolumn{3}{c}{F1-Score$\uparrow$}     & \multicolumn{3}{c}{NC$\uparrow$}           \\ 
   & SuperDec~\cite{fedele2025superdec} & MP~\cite{liu2023marching}    & our method     & SuperDec~\cite{fedele2025superdec}   & MP~\cite{liu2023marching}    & our method \\ \hline
bench    & 0.8659  & 0.8823   & \textbf{0.8434}       & 0.7720  & 0.4789  & \textbf{0.3542}     \\ 
table    & 0.8790  & 0.8862   & \textbf{0.8913}        & 0.8207 & 0.4174  & \textbf{0.4061}      \\ 
plane    & 0.8779 & 0.8760 & \textbf{0.9046}        & 0.6571 & 0.6029 & \textbf{0.5328}      \\ 
cabinet  & 0.8596 & 0.8818 & \textbf{0.8942}        & 0.7841 & 0.5245  & \textbf{0.4430}      \\
bottle   & 0.9071  & 0.8556  & \textbf{0.8918}        & 0.8925 & 0.5067  & \textbf{0.4641}      \\ 
rifle    & 0.7464 & 0.8772  & \textbf{0.9122}        & 0.6799  & 0.5456  & \textbf{0.5067}     \\ \hline
mean     & 0.8560  & 0.8765  & \textbf{0.8896}        & 0.7677 & 0.5127 & \textbf{0.4511}      \\ 
\hline

\end{tabular}
\end{table*}

We selected five types of models on the ShapeNet in virtual scene dataset for comparative experiments, to further validate the performance of 3D model generation in virtual scenes. The quantitative and qualitative experiments on the ShapNet dataset are shown in Tab~\ref{tab_sha} and Fig~\ref{fig_sha}, respectively. From the table, we can see that our method performs best in various categories and indicators, achieving an average CD of $0.503\times10^{-3}$, a VIoU of 0.742, a F1-Score of 0.8896, and a NC of 0.4511. From the figure, we can see that our method has similar results to MP's generation of bench and table models. For the generation of the other three models, our method has a smoother surface. This is because on the other three models, our method uses ellipsoids with smooth surfaces to optimize the superquadric with lateral edges. The surface quality of the generated model is improved, which also verifies that our method has a certain generalization performance.

\subsubsection{Validation experiment of primitive fitting effect on real scene dataset}

\begin{figure}[ht]
  \centering
  \includegraphics[width=0.6\linewidth]{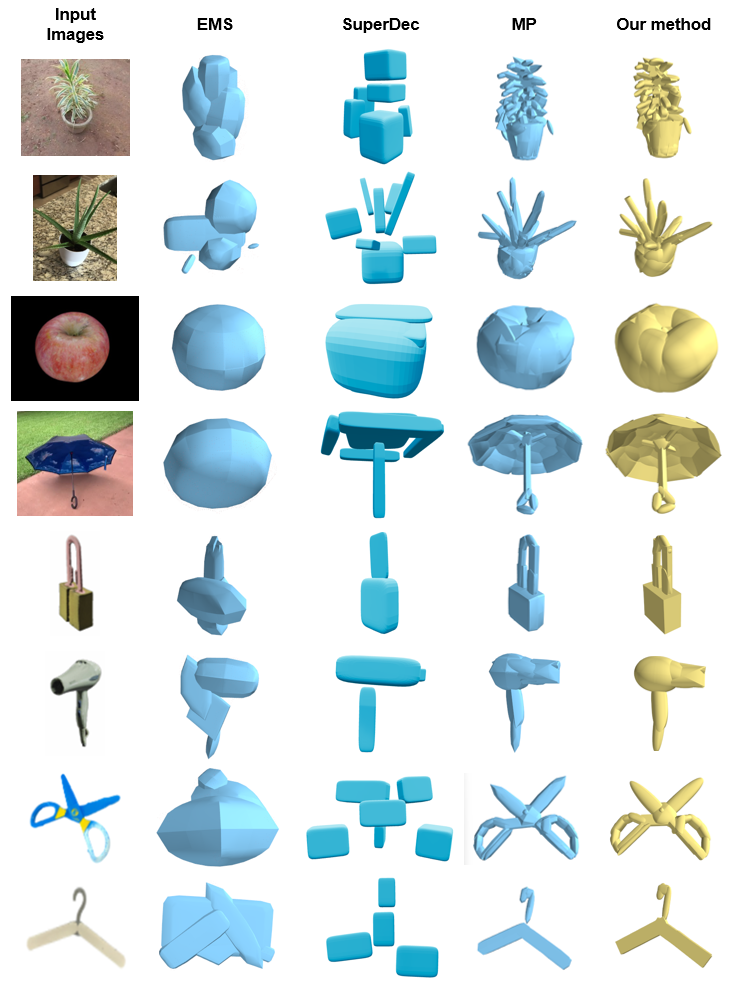}
  \caption{Qualitative experimental results on real scene dataset. The first four rows of input images have background information, while the rest of the input images do not have background information}\label{fig_real}
\end{figure}

We selected different objects in real scene to verify the effectiveness of the model generation. The qualitative and quantitative experimental results on the real scene dataset are shown in Fig~\ref{fig_real} and tab~\ref{tab_real}, respectively. From the qualitative results, it can be seen that our method can effectively generate target models with zero-shot under both background and no background conditions, and the model has smoother surface quality and a shape closer to real objects. From the quantitative results, it can be seen that our method achieves the best in all four indicators, further demonstrating the generalization of our method in real-world scenarios.

\subsubsection{Comparative experiment on storage capacity of generated models}

    One of the goals of our method is to reduce the storage requirements for generating models. In order to verify the effectiveness of our storage method, we counted the average storage capacity of 60 generation models directly extracted into Obj files through the Marching Cubes algorithm through experiments, and the corresponding storage capacity of primitive parameter files after executing our method. Our model storage method is introduced in~\ref{sec1}. The experimental results can be seen in Table~\ref{tab3}. It can be seen that the storage size is reduced by three orders of magnitude. This is because the parametric representation of primitives can efficiently describe complex 3D shapes with a few parameters, and eliminate unnecessary details through simplification and abstraction. In contrast, the mesh model needs more storage space to save all the information of the model due to its detailed geometric representation and complex storage format. Therefore, the reduction of the storage capacity of the generation model is of great significance for the large number of applications of the generation model in the graphics system.

    \begin{table}[h]
    \centering
    \caption{Quantitative experimental results on the real scene dataset}
    \label{tab_real}
    \begin{tabular}{ccccc}
    \hline
         & EMS~\cite{liu2022robust} & SuperDec~\cite{fedele2025superdec}   & MP~\cite{liu2023marching}    & our method \\ \hline
    CD($\times10^{-3}$)$\downarrow$ & 15.1 & 4.40 & 4.32  & \textbf{2.52}      \\
    VIoU$\uparrow$     & 0.141 & 0.301 & 0.492 & \textbf{0.673}      \\ 
    F1-Score$\uparrow$     & 0.8917 & 0.8383 & 0.7771   & \textbf{0.9183}      \\ 
    NC$\uparrow$     & 0.7539 & 0.6759 & 0.5882  & \textbf{0.7752}      \\ \hline
    \end{tabular}
    \end{table}

\begin{table}[ht]
 \caption{Generation model storage capacity comparison experiment.}
 \centering
\begin{tabular}{ccc}
\hline
Input type & \begin{tabular}[c]{@{}c@{}}Mesh \\ storage capacity\end{tabular} & \begin{tabular}[c]{@{}c@{}}Primitive \\ storage capacity\end{tabular} \\ \hline
Texts & 4.56MB & 5KB \\ 
\begin{tabular}[c]{@{}c@{}}Images\end{tabular} & 5.76MB & 6KB \\ 
All & 5.36MB & 6KB \\ \hline
\end{tabular}\label{tab3}
\end{table}

\subsubsection{Ablation study}
To further validate the effectiveness of the primitive body fitting and matching algorithm, we conducted ablation studies and compared our algorithm with several other variants. The experiments were conducted on the real scene dataset. From Fig~\ref{figequ}, it can be seen that parameterized geometry is represented by four polar coordinate equations (1, 2, 3 and 5). Therefore, we design four variants based on these equations, called primitive fitting and matching variants x (PFMx, $x=1,2,3,4$). In each variant, the generated target 3D model is based solely on one parameterized geometry. For example, in PFM1, the parameterized geometry is only represented by equation 1 in Fig~\ref{figequ}, in PFM3, the parameterized geometry is only represented by equation 4 in Fig~\ref{figequ}. In addition to these variants, MP~\cite{liu2023marching} was also added for comparison.

\begin{table}[h]
\centering
\caption{Ablation study results on the real scene dataset}
\label{tab_abla}
\begin{tabular}{ccccccc}
\hline
& MP~\cite{liu2023marching}     & PFM1  & PFM2 & PFM3 & PFM4 & our method \\ \hline
CD($\times10^{-3}$)$\downarrow$       & 4.37   & 2.50    & 2.38 & 2.33 & 2.43& \textbf{2.41}       \\
VIoU$\uparrow$     & 0.460  & 0.527   & 0.586 & 0.608 & 0.663& \textbf{0.681}      \\
F1-Score$\uparrow$ & 0.7763 & 0.02800 & 0.04940 & 0.03111 & 0.08396& \textbf{0.9247} \\
NC$\uparrow$        & 0.5713 & 0.7132  & 0.7488 & 0.7286 & 0.7749& \textbf{0.7991}  \\ \hline  
\end{tabular}\label{tab4}
\end{table}

The ablation study results on the real scene dataset are shown in Table~\ref{tab4}. It can be observed that although our method do not perform the best on CD, it still perform best on VIoU, F1-Score and NC. This is because our method takes into account the shape parameters of the superquadrics, $\varepsilon_1$ and $\varepsilon_2$, which makes our method more adaptable to the shape of the superquadrics than the variants, thus generating a 3D model closer to the target model. The experimental results also prove the effectiveness of our four polar coordinate equations.

\subsection{The limitations and deficiencies of the algorithm}

    Our method realizes a parameterized primitives synthesis model based on cross modality information. It has achieved certain effects in simple model testing. However, there are still the following limitations and deficiencies. First, in terms of primitive matching, our method cannot achieve effective matching or fitting for toroidal column. This is because the superquadric has no penetrating surface. Second, During the experimental process, we were unable to demonstrate the advantages of our parameterized representation over other alternatives such as NURBS. Finally, our method is limited by the quality of multi-view generation, and the model quality of some invisible perspectives of complex models is limited.

    For these problems, we also made attempts at solutions. For the primitive matching of toroidal column, we tried to use variational autoencoders to encode the point clouds of complex primitive bodies and use the point cloud encoding features for matching. For the demonstration of the advantages of parameterized representation problem, we tried to use other types of surfaces to fit the components of the model. For the model quality problem of invisible perspectives, we tried to simultaneously utilize different modalities information to better describe the features of the target model, or perform fine-tuning training in downstream tasks to improve the quality of multi-view generation. However, due to time limitations, no results have been achieved yet. We hope this can bring some inspiration.

\section{Conclusion}

This paper proposes a multi-stage method based on cross modality to generate parameterized primitive combination model. A novel parameterized primitive synthesis algorithm and a model storage method are proposed in the frame. The experimental result demonstrates that our method is capable of generating diverse 3D basic models in response to a wide range of conditional inputs, and outperforms state-of-the-art algorithms in terms of CD, VIoU, F1-Score and NC on both virtual scene and real scene datasets. It also demonstrates that parameterized primitive synthesis model is more in line with aesthetic requirements. However, our method still facing the problem of the fitting of toroidal column, demonstration of the advantages of parameterized representation and the invisible perspectives perspectives. In the future, our work will focus on variational autoencoder, parameterized primitive fitting based on other surfaces, and better describing the features of the target model.
\\

\bibliographystyle{elsarticle-num-names.bst}
\bibliography{reference.bib}

\begin{thebibliography}{38}
\expandafter\ifx\csname natexlab\endcsname\relax\def\natexlab#1{#1}\fi
\providecommand{\url}[1]{\texttt{#1}}
\providecommand{\href}[2]{#2}
\providecommand{\path}[1]{#1}
\providecommand{\DOIprefix}{doi:}
\providecommand{\ArXivprefix}{arXiv:}
\providecommand{\URLprefix}{URL: }
\providecommand{\Pubmedprefix}{pmid:}
\providecommand{\doi}[1]{\href{http://dx.doi.org/#1}{\path{#1}}}
\providecommand{\Pubmed}[1]{\href{pmid:#1}{\path{#1}}}
\providecommand{\bibinfo}[2]{#2}
\ifx\xfnm\relax \def\xfnm[#1]{\unskip,\space#1}\fi
\bibitem[{Li et~al.(2023)Li, Zhang, Cho, Waghwase, Lee, Rameau, Yang, Bae, and Hong}]{li2023generative}
\bibinfo{author}{C.~Li}, \bibinfo{author}{C.~Zhang}, \bibinfo{author}{J.~Cho}, \bibinfo{author}{A.~Waghwase}, \bibinfo{author}{L.-H. Lee}, \bibinfo{author}{F.~Rameau}, \bibinfo{author}{Y.~Yang}, \bibinfo{author}{S.-H. Bae}, \bibinfo{author}{C.~S. Hong},
\newblock \bibinfo{title}{Generative ai meets 3d: A survey on text-to-3d in aigc era},
\newblock \bibinfo{journal}{arXiv preprint arXiv:2305.06131}  (\bibinfo{year}{2023}).
\bibitem[{Zi et~al.(2017)Zi, Cong, and Zhang}]{Zilingling2017Survey}
\bibinfo{author}{L.~Zi}, \bibinfo{author}{X.~Cong}, \bibinfo{author}{Y.~Zhang},
\newblock \bibinfo{title}{Survey on semantics driven 3d model creation},
\newblock \bibinfo{journal}{Application Research of Computers} \bibinfo{volume}{34} (\bibinfo{year}{2017}) \bibinfo{pages}{641--646}.
\bibitem[{Furukawa et~al.(2010)Furukawa, Curless, Seitz, and Szeliski}]{furukawa2010towards}
\bibinfo{author}{Y.~Furukawa}, \bibinfo{author}{B.~Curless}, \bibinfo{author}{S.~M. Seitz}, \bibinfo{author}{R.~Szeliski},
\newblock \bibinfo{title}{Towards internet-scale multi-view stereo},
\newblock in: \bibinfo{booktitle}{2010 IEEE computer society conference on computer vision and pattern recognition}, \bibinfo{organization}{IEEE}, \bibinfo{year}{2010}, pp. \bibinfo{pages}{1434--1441}.
\bibitem[{Sch{\"o}nberger et~al.(2016)Sch{\"o}nberger, Zheng, Frahm, and Pollefeys}]{schonberger2016pixelwise}
\bibinfo{author}{J.~L. Sch{\"o}nberger}, \bibinfo{author}{E.~Zheng}, \bibinfo{author}{J.-M. Frahm}, \bibinfo{author}{M.~Pollefeys},
\newblock \bibinfo{title}{Pixelwise view selection for unstructured multi-view stereo},
\newblock in: \bibinfo{booktitle}{Computer Vision--ECCV 2016: 14th European Conference, Amsterdam, The Netherlands, October 11-14, 2016, Proceedings, Part III 14}, \bibinfo{organization}{Springer}, \bibinfo{year}{2016}, pp. \bibinfo{pages}{501--518}.
\bibitem[{Choi et~al.(2010)Choi, Medioni, Lin, Silva, Regina, Pamplona, and Faltemier}]{choi20103d}
\bibinfo{author}{J.~Choi}, \bibinfo{author}{G.~Medioni}, \bibinfo{author}{Y.~Lin}, \bibinfo{author}{L.~Silva}, \bibinfo{author}{O.~Regina}, \bibinfo{author}{M.~Pamplona}, \bibinfo{author}{T.~C. Faltemier},
\newblock \bibinfo{title}{3d face reconstruction using a single or multiple views},
\newblock in: \bibinfo{booktitle}{2010 20th International Conference on Pattern Recognition}, \bibinfo{organization}{IEEE}, \bibinfo{year}{2010}, pp. \bibinfo{pages}{3959--3962}.
\bibitem[{Min et~al.(2023)Min, Cha, Won, and Lim}]{min2023tsdf}
\bibinfo{author}{C.~Min}, \bibinfo{author}{S.~Cha}, \bibinfo{author}{C.~Won}, \bibinfo{author}{J.~Lim},
\newblock \bibinfo{title}{Tsdf-sampling: Efficient sampling for neural surface field using truncated signed distance field},
\newblock \bibinfo{journal}{arXiv preprint arXiv:2311.17878}  (\bibinfo{year}{2023}).
\bibitem[{Tatarchenko et~al.(2017)Tatarchenko, Dosovitskiy, and Brox}]{tatarchenko2017octree}
\bibinfo{author}{M.~Tatarchenko}, \bibinfo{author}{A.~Dosovitskiy}, \bibinfo{author}{T.~Brox},
\newblock \bibinfo{title}{Octree generating networks: Efficient convolutional architectures for high-resolution 3d outputs},
\newblock in: \bibinfo{booktitle}{Proceedings of the IEEE international conference on computer vision}, \bibinfo{year}{2017}, pp. \bibinfo{pages}{2088--2096}.
\bibitem[{Lorensen(2020)}]{lorensen2020history}
\bibinfo{author}{W.~E. Lorensen},
\newblock \bibinfo{title}{History of the marching cubes algorithm},
\newblock \bibinfo{journal}{IEEE computer graphics and applications} \bibinfo{volume}{40} (\bibinfo{year}{2020}) \bibinfo{pages}{8--15}.
\bibitem[{Mildenhall et~al.(2021)Mildenhall, Srinivasan, Tancik, Barron, Ramamoorthi, and Ng}]{mildenhall2021nerf}
\bibinfo{author}{B.~Mildenhall}, \bibinfo{author}{P.~P. Srinivasan}, \bibinfo{author}{M.~Tancik}, \bibinfo{author}{J.~T. Barron}, \bibinfo{author}{R.~Ramamoorthi}, \bibinfo{author}{R.~Ng},
\newblock \bibinfo{title}{Nerf: Representing scenes as neural radiance fields for view synthesis},
\newblock \bibinfo{journal}{Communications of the ACM} \bibinfo{volume}{65} (\bibinfo{year}{2021}) \bibinfo{pages}{99--106}.
\bibitem[{Kerbl et~al.(2023)Kerbl, Kopanas, Leimk{\"u}hler, and Drettakis}]{kerbl20233d}
\bibinfo{author}{B.~Kerbl}, \bibinfo{author}{G.~Kopanas}, \bibinfo{author}{T.~Leimk{\"u}hler}, \bibinfo{author}{G.~Drettakis},
\newblock \bibinfo{title}{3d gaussian splatting for real-time radiance field rendering},
\newblock \bibinfo{journal}{ACM Transactions on Graphics} \bibinfo{volume}{42} (\bibinfo{year}{2023}) \bibinfo{pages}{1--14}.
\bibitem[{Golyanik et~al.(2018)Golyanik, Shimada, Varanasi, and Stricker}]{golyanik2018hdm}
\bibinfo{author}{V.~Golyanik}, \bibinfo{author}{S.~Shimada}, \bibinfo{author}{K.~Varanasi}, \bibinfo{author}{D.~Stricker},
\newblock \bibinfo{title}{Hdm-net: Monocular non-rigid 3d reconstruction with learned deformation model},
\newblock in: \bibinfo{booktitle}{Virtual Reality and Augmented Reality: 15th EuroVR International Conference, EuroVR 2018, London, UK, October 22--23, 2018, Proceedings 15}, \bibinfo{organization}{Springer}, \bibinfo{year}{2018}, pp. \bibinfo{pages}{51--72}.
\bibitem[{Bednarik et~al.(2018)Bednarik, Fua, and Salzmann}]{bednarik2018learning}
\bibinfo{author}{J.~Bednarik}, \bibinfo{author}{P.~Fua}, \bibinfo{author}{M.~Salzmann},
\newblock \bibinfo{title}{Learning to reconstruct texture-less deformable surfaces from a single view},
\newblock in: \bibinfo{booktitle}{2018 international conference on 3d vision (3DV)}, \bibinfo{organization}{IEEE}, \bibinfo{year}{2018}, pp. \bibinfo{pages}{606--615}.
\bibitem[{Wu et~al.(2017)Wu, Wang, Xue, Sun, Freeman, and Tenenbaum}]{wu2017marrnet}
\bibinfo{author}{J.~Wu}, \bibinfo{author}{Y.~Wang}, \bibinfo{author}{T.~Xue}, \bibinfo{author}{X.~Sun}, \bibinfo{author}{B.~Freeman}, \bibinfo{author}{J.~Tenenbaum},
\newblock \bibinfo{title}{Marrnet: 3d shape reconstruction via 2.5 d sketches},
\newblock \bibinfo{journal}{Advances in neural information processing systems} \bibinfo{volume}{30} (\bibinfo{year}{2017}).
\bibitem[{Choy et~al.(2016)Choy, Xu, Gwak, Chen, and Savarese}]{choy20163d}
\bibinfo{author}{C.~B. Choy}, \bibinfo{author}{D.~Xu}, \bibinfo{author}{J.~Gwak}, \bibinfo{author}{K.~Chen}, \bibinfo{author}{S.~Savarese},
\newblock \bibinfo{title}{3d-r2n2: A unified approach for single and multi-view 3d object reconstruction},
\newblock in: \bibinfo{booktitle}{Computer Vision--ECCV 2016: 14th European Conference, Amsterdam, The Netherlands, October 11-14, 2016, Proceedings, Part VIII 14}, \bibinfo{organization}{Springer}, \bibinfo{year}{2016}, pp. \bibinfo{pages}{628--644}.
\bibitem[{Kar et~al.(2017)Kar, H{\"a}ne, and Malik}]{kar2017learning}
\bibinfo{author}{A.~Kar}, \bibinfo{author}{C.~H{\"a}ne}, \bibinfo{author}{J.~Malik},
\newblock \bibinfo{title}{Learning a multi-view stereo machine},
\newblock \bibinfo{journal}{Advances in neural information processing systems} \bibinfo{volume}{30} (\bibinfo{year}{2017}).
\bibitem[{Ho et~al.(2020)Ho, Jain, and Abbeel}]{ho2020denoising}
\bibinfo{author}{J.~Ho}, \bibinfo{author}{A.~Jain}, \bibinfo{author}{P.~Abbeel},
\newblock \bibinfo{title}{Denoising diffusion probabilistic models},
\newblock \bibinfo{journal}{Advances in neural information processing systems} \bibinfo{volume}{33} (\bibinfo{year}{2020}) \bibinfo{pages}{6840--6851}.
\bibitem[{Wang and Shi(2023)}]{wang2023imagedream}
\bibinfo{author}{P.~Wang}, \bibinfo{author}{Y.~Shi},
\newblock \bibinfo{title}{Imagedream: Image-prompt multi-view diffusion for 3d generation},
\newblock \bibinfo{journal}{arXiv preprint arXiv:2312.02201}  (\bibinfo{year}{2023}).
\bibitem[{Shi et~al.(2023)Shi, Wang, Ye, Mai, Li, and Yang}]{shi2023mvdream}
\bibinfo{author}{Y.~Shi}, \bibinfo{author}{P.~Wang}, \bibinfo{author}{J.~Ye}, \bibinfo{author}{L.~Mai}, \bibinfo{author}{K.~Li}, \bibinfo{author}{X.~Yang},
\newblock \bibinfo{title}{Mvdream: Multi-view diffusion for 3d generation},
\newblock in: \bibinfo{booktitle}{The Twelfth International Conference on Learning Representations}, \bibinfo{year}{2023}.
\bibitem[{Long et~al.(2023)Long, Guo, Lin, Liu, Dou, Liu, Ma, Zhang, Habermann, Theobalt et~al.}]{long2023wonder3d}
\bibinfo{author}{X.~Long}, \bibinfo{author}{Y.-C. Guo}, \bibinfo{author}{C.~Lin}, \bibinfo{author}{Y.~Liu}, \bibinfo{author}{Z.~Dou}, \bibinfo{author}{L.~Liu}, \bibinfo{author}{Y.~Ma}, \bibinfo{author}{S.-H. Zhang}, \bibinfo{author}{M.~Habermann}, \bibinfo{author}{C.~Theobalt}, et~al.,
\newblock \bibinfo{title}{Wonder3d: Single image to 3d using cross-domain diffusion},
\newblock \bibinfo{journal}{arXiv preprint arXiv:2310.15008}  (\bibinfo{year}{2023}).
\bibitem[{Liu et~al.(2023)Liu, Wu, Van~Hoorick, Tokmakov, Zakharov, and Vondrick}]{liu2023zero}
\bibinfo{author}{R.~Liu}, \bibinfo{author}{R.~Wu}, \bibinfo{author}{B.~Van~Hoorick}, \bibinfo{author}{P.~Tokmakov}, \bibinfo{author}{S.~Zakharov}, \bibinfo{author}{C.~Vondrick},
\newblock \bibinfo{title}{Zero-1-to-3: Zero-shot one image to 3d object},
\newblock in: \bibinfo{booktitle}{Proceedings of the IEEE/CVF International Conference on Computer Vision}, \bibinfo{year}{2023}, pp. \bibinfo{pages}{9298--9309}.
\bibitem[{Abdullah et~al.(2024)Abdullah, Rahman, Rahman, and Islam}]{abdullah2024vae}
\bibinfo{author}{M.~T. Abdullah}, \bibinfo{author}{S.~Rahman}, \bibinfo{author}{S.~Rahman}, \bibinfo{author}{M.~F. Islam},
\newblock \bibinfo{title}{Vae-gan3d: Leveraging image-based semantics for 3d zero-shot recognition},
\newblock \bibinfo{journal}{Image and Vision Computing} \bibinfo{volume}{147} (\bibinfo{year}{2024}) \bibinfo{pages}{105049}.
\bibitem[{Kluger et~al.(2021)Kluger, Ackermann, Brachmann, Yang, and Rosenhahn}]{kluger2021cuboids}
\bibinfo{author}{F.~Kluger}, \bibinfo{author}{H.~Ackermann}, \bibinfo{author}{E.~Brachmann}, \bibinfo{author}{M.~Y. Yang}, \bibinfo{author}{B.~Rosenhahn},
\newblock \bibinfo{title}{Cuboids revisited: Learning robust 3d shape fitting to single rgb images},
\newblock in: \bibinfo{booktitle}{Proceedings of the IEEE/CVF Conference on Computer Vision and Pattern Recognition}, \bibinfo{year}{2021}, pp. \bibinfo{pages}{13070--13079}.
\bibitem[{Genova et~al.(2019)Genova, Cole, Vlasic, Sarna, Freeman, and Funkhouser}]{genova2019learning}
\bibinfo{author}{K.~Genova}, \bibinfo{author}{F.~Cole}, \bibinfo{author}{D.~Vlasic}, \bibinfo{author}{A.~Sarna}, \bibinfo{author}{W.~T. Freeman}, \bibinfo{author}{T.~Funkhouser},
\newblock \bibinfo{title}{Learning shape templates with structured implicit functions},
\newblock in: \bibinfo{booktitle}{Proceedings of the IEEE/CVF international conference on computer vision}, \bibinfo{year}{2019}, pp. \bibinfo{pages}{7154--7164}.
\bibitem[{Fahim et~al.(2022)Fahim, Amin, and Zarif}]{fahim2022enhancing}
\bibinfo{author}{G.~Fahim}, \bibinfo{author}{K.~Amin}, \bibinfo{author}{S.~Zarif},
\newblock \bibinfo{title}{Enhancing single-view 3d mesh reconstruction with the aid of implicit surface learning},
\newblock \bibinfo{journal}{Image and Vision Computing} \bibinfo{volume}{119} (\bibinfo{year}{2022}) \bibinfo{pages}{104377}.
\bibitem[{Smirnov et~al.(2020)Smirnov, Fisher, Kim, Zhang, and Solomon}]{smirnov2020deep}
\bibinfo{author}{D.~Smirnov}, \bibinfo{author}{M.~Fisher}, \bibinfo{author}{V.~G. Kim}, \bibinfo{author}{R.~Zhang}, \bibinfo{author}{J.~Solomon},
\newblock \bibinfo{title}{Deep parametric shape predictions using distance fields},
\newblock in: \bibinfo{booktitle}{Proceedings of the IEEE/CVF Conference on Computer Vision and Pattern Recognition}, \bibinfo{year}{2020}, pp. \bibinfo{pages}{561--570}.
\bibitem[{Paschalidou et~al.(2021)Paschalidou, Katharopoulos, Geiger, and Fidler}]{paschalidou2021neural}
\bibinfo{author}{D.~Paschalidou}, \bibinfo{author}{A.~Katharopoulos}, \bibinfo{author}{A.~Geiger}, \bibinfo{author}{S.~Fidler},
\newblock \bibinfo{title}{Neural parts: Learning expressive 3d shape abstractions with invertible neural networks},
\newblock in: \bibinfo{booktitle}{Proceedings of the IEEE/CVF Conference on Computer Vision and Pattern Recognition}, \bibinfo{year}{2021}, pp. \bibinfo{pages}{3204--3215}.
\bibitem[{Saporta and Sharf(2022)}]{saporta2022unsupervised}
\bibinfo{author}{T.~Saporta}, \bibinfo{author}{A.~Sharf},
\newblock \bibinfo{title}{Unsupervised recursive deep fitting of 3d primitives to points},
\newblock \bibinfo{journal}{Computers \& Graphics} \bibinfo{volume}{102} (\bibinfo{year}{2022}) \bibinfo{pages}{289--299}.
\bibitem[{Liu et~al.(2023)Liu, Wu, Ruan, and Chirikjian}]{liu2023marching}
\bibinfo{author}{W.~Liu}, \bibinfo{author}{Y.~Wu}, \bibinfo{author}{S.~Ruan}, \bibinfo{author}{G.~S. Chirikjian},
\newblock \bibinfo{title}{Marching-primitives: Shape abstraction from signed distance function},
\newblock in: \bibinfo{booktitle}{Proceedings of the IEEE/CVF Conference on Computer Vision and Pattern Recognition}, \bibinfo{year}{2023}, pp. \bibinfo{pages}{8771--8780}.
\bibitem[{Poole et~al.(2022)Poole, Jain, Barron, and Mildenhall}]{poole2022dreamfusion}
\bibinfo{author}{B.~Poole}, \bibinfo{author}{A.~Jain}, \bibinfo{author}{J.~T. Barron}, \bibinfo{author}{B.~Mildenhall},
\newblock \bibinfo{title}{Dreamfusion: Text-to-3d using 2d diffusion},
\newblock in: \bibinfo{booktitle}{The Eleventh International Conference on Learning Representations}, \bibinfo{year}{2022}.
\bibitem[{Tancik et~al.(2023)Tancik, Weber, Ng, Li, Yi, Wang, Kristoffersen, Austin, Salahi, Ahuja et~al.}]{tancik2023nerfstudio}
\bibinfo{author}{M.~Tancik}, \bibinfo{author}{E.~Weber}, \bibinfo{author}{E.~Ng}, \bibinfo{author}{R.~Li}, \bibinfo{author}{B.~Yi}, \bibinfo{author}{T.~Wang}, \bibinfo{author}{A.~Kristoffersen}, \bibinfo{author}{J.~Austin}, \bibinfo{author}{K.~Salahi}, \bibinfo{author}{A.~Ahuja}, et~al.,
\newblock \bibinfo{title}{Nerfstudio: A modular framework for neural radiance field development},
\newblock in: \bibinfo{booktitle}{ACM SIGGRAPH 2023 Conference Proceedings}, \bibinfo{year}{2023}, pp. \bibinfo{pages}{1--12}.
\bibitem[{Chang et~al.(2015)Chang, Funkhouser, Guibas, Hanrahan, Huang, Li, Savarese, Savva, Song, Su et~al.}]{chang2015shapenet}
\bibinfo{author}{A.~X. Chang}, \bibinfo{author}{T.~Funkhouser}, \bibinfo{author}{L.~Guibas}, \bibinfo{author}{P.~Hanrahan}, \bibinfo{author}{Q.~Huang}, \bibinfo{author}{Z.~Li}, \bibinfo{author}{S.~Savarese}, \bibinfo{author}{M.~Savva}, \bibinfo{author}{S.~Song}, \bibinfo{author}{H.~Su}, et~al.,
\newblock \bibinfo{title}{Shapenet: An information-rich 3d model repository},
\newblock \bibinfo{journal}{arXiv preprint arXiv:1512.03012}  (\bibinfo{year}{2015}).
\bibitem[{Liu et~al.(2024)Liu, Shi, Chen, Zhang, Xu, Wei, Chen, Zeng, Gu, and Su}]{liu2024one}
\bibinfo{author}{M.~Liu}, \bibinfo{author}{R.~Shi}, \bibinfo{author}{L.~Chen}, \bibinfo{author}{Z.~Zhang}, \bibinfo{author}{C.~Xu}, \bibinfo{author}{X.~Wei}, \bibinfo{author}{H.~Chen}, \bibinfo{author}{C.~Zeng}, \bibinfo{author}{J.~Gu}, \bibinfo{author}{H.~Su},
\newblock \bibinfo{title}{One-2-3-45++: Fast single image to 3d objects with consistent multi-view generation and 3d diffusion},
\newblock in: \bibinfo{booktitle}{Proceedings of the IEEE/CVF conference on computer vision and pattern recognition}, \bibinfo{year}{2024}, pp. \bibinfo{pages}{10072--10083}.
\bibitem[{Tochilkin et~al.(2024)Tochilkin, Pankratz, Liu, Huang, Letts, Li, Liang, Laforte, Jampani, and Cao}]{tochilkin2024triposr}
\bibinfo{author}{D.~Tochilkin}, \bibinfo{author}{D.~Pankratz}, \bibinfo{author}{Z.~Liu}, \bibinfo{author}{Z.~Huang}, \bibinfo{author}{A.~Letts}, \bibinfo{author}{Y.~Li}, \bibinfo{author}{D.~Liang}, \bibinfo{author}{C.~Laforte}, \bibinfo{author}{V.~Jampani}, \bibinfo{author}{Y.-P. Cao},
\newblock \bibinfo{title}{Triposr: Fast 3d object reconstruction from a single image},
\newblock \bibinfo{journal}{arXiv preprint arXiv:2403.02151}  (\bibinfo{year}{2024}).
\bibitem[{Reizenstein et~al.(2021)Reizenstein, Shapovalov, Henzler, Sbordone, Labatut, and Novotny}]{reizenstein2021common}
\bibinfo{author}{J.~Reizenstein}, \bibinfo{author}{R.~Shapovalov}, \bibinfo{author}{P.~Henzler}, \bibinfo{author}{L.~Sbordone}, \bibinfo{author}{P.~Labatut}, \bibinfo{author}{D.~Novotny},
\newblock \bibinfo{title}{Common objects in 3d: Large-scale learning and evaluation of real-life 3d category reconstruction},
\newblock in: \bibinfo{booktitle}{Proceedings of the IEEE/CVF international conference on computer vision}, \bibinfo{year}{2021}, pp. \bibinfo{pages}{10901--10911}.
\bibitem[{Liu et~al.(2022)Liu, Xu, Fu, Qian, Yu, Han, and Lu}]{liu2022akb}
\bibinfo{author}{L.~Liu}, \bibinfo{author}{W.~Xu}, \bibinfo{author}{H.~Fu}, \bibinfo{author}{S.~Qian}, \bibinfo{author}{Q.~Yu}, \bibinfo{author}{Y.~Han}, \bibinfo{author}{C.~Lu},
\newblock \bibinfo{title}{Akb-48: A real-world articulated object knowledge base},
\newblock in: \bibinfo{booktitle}{Proceedings of the IEEE/CVF Conference on Computer Vision and Pattern Recognition}, \bibinfo{year}{2022}, pp. \bibinfo{pages}{14809--14818}.
\bibitem[{Wu et~al.(2023)Wu, Zhang, Fu, Wang, Ren, Pan, Wu, Yang, Wang, Qian et~al.}]{wu2023omniobject3d}
\bibinfo{author}{T.~Wu}, \bibinfo{author}{J.~Zhang}, \bibinfo{author}{X.~Fu}, \bibinfo{author}{Y.~Wang}, \bibinfo{author}{J.~Ren}, \bibinfo{author}{L.~Pan}, \bibinfo{author}{W.~Wu}, \bibinfo{author}{L.~Yang}, \bibinfo{author}{J.~Wang}, \bibinfo{author}{C.~Qian}, et~al.,
\newblock \bibinfo{title}{Omniobject3d: Large-vocabulary 3d object dataset for realistic perception, reconstruction and generation},
\newblock in: \bibinfo{booktitle}{Proceedings of the IEEE/CVF Conference on Computer Vision and Pattern Recognition}, \bibinfo{year}{2023}, pp. \bibinfo{pages}{803--814}.
\bibitem[{Liu et~al.(2022)Liu, Wu, Ruan, and Chirikjian}]{liu2022robust}
\bibinfo{author}{W.~Liu}, \bibinfo{author}{Y.~Wu}, \bibinfo{author}{S.~Ruan}, \bibinfo{author}{G.~S. Chirikjian},
\newblock \bibinfo{title}{Robust and accurate superquadric recovery: A probabilistic approach},
\newblock in: \bibinfo{booktitle}{Proceedings of the IEEE/CVF Conference on Computer Vision and Pattern Recognition}, \bibinfo{year}{2022}, pp. \bibinfo{pages}{2676--2685}.
\bibitem[{Fedele et~al.(2025)Fedele, Sun, Guibas, Pollefeys, and Engelmann}]{fedele2025superdec}
\bibinfo{author}{E.~Fedele}, \bibinfo{author}{B.~Sun}, \bibinfo{author}{L.~Guibas}, \bibinfo{author}{M.~Pollefeys}, \bibinfo{author}{F.~Engelmann},
\newblock \bibinfo{title}{Superdec: 3d scene decomposition with superquadric primitives},
\newblock \bibinfo{journal}{arXiv preprint arXiv:2504.00992}  (\bibinfo{year}{2025}).

\end{thebibliography}

\end{document}